\documentclass[apj]{emulateapj}

\usepackage{times}

\shorttitle{Neon abundance in the Sun-as-a-star}                    
\shortauthors{Brooks}

\begin{document}

%% ------------------------------------------------------------------------------------------
%% --- TITLE PAGE ---------------------------------------------------------------------------
%% ------------------------------------------------------------------------------------------

\title{Solar cycle observations of the Neon abundance in the Sun-as-a-star}

\author{David H. Brooks\altaffilmark{1,2} and Deborah Baker\altaffilmark{3}, Lidia van Driel-Gesztelyi \altaffilmark{3,4, 5} and Harry P. Warren\altaffilmark{6}}

\affiliation{\altaffilmark{1}College of Science, George Mason University, 4400 University Drive,
  Fairfax, VA 22030}

\affiliation{\altaffilmark{3}Mullard Space Science Laboratory, University College London, Holmbury St. Mary, Dorking, Surrey RH5 6NT, UK}

\affiliation{\altaffilmark{4}Observatoire de Paris, LESIA, UMR 8109 (CNRS), F-92195 Meudon Principal Cedex, France}

\affiliation{\altaffilmark{5}Konkoly Observatory of the Hungarian Academy of Sciences, Budapest, Hungary}

\affiliation{\altaffilmark{6}Space Science Division, Naval Research Laboratory, Washington, DC 20375, USA}

\altaffiltext{2}{Current address: Hinode Team, ISAS/JAXA, 3-1-1 Yoshinodai, Chuo-ku, Sagamihara,
  Kanagawa 252-5210, Japan}

%% ------------------------------------------------------------------------------------------
%% --- ABSTRACT -----------------------------------------------------------------------------
%% ------------------------------------------------------------------------------------------

\begin{abstract}
Properties of the Sun's interior can be determined accurately
from helioseismological measurements of solar oscillations. These measurements, 
however, are in conflict with photospheric elemental abundances derived using 3-D hydrodynamic models of the solar atmosphere. 
This divergence of theory and helioseismology is known as the $``$solar modeling problem$"$.
One possible solution is that the photospheric neon abundance, which is deduced indirectly 
by combining the coronal Ne/O ratio with the photospheric O abundance, is larger
than generally accepted. 
There is some support for this idea from observations of cool stars. 
The Ne/O abundance
ratio has also been found to vary with the solar cycle in the slowest solar wind streams and coronal
streamers, and the variation from solar maximum to minimum in streamers ($\sim$0.1 to 0.25) is large enough
to potentially bring some of the solar models
into agreement with the seismic data.
Here we use daily-sampled observations from the EUV Variability Experiment (EVE) on the Solar Dynamics Observatory taken in 2010--2014,
to investigate whether the coronal Ne/O abundance ratio shows a variation with the solar cycle when the Sun
is viewed as a star. We find only a weak dependence on, and moderate anti-correlation with, the solar cycle
with the ratio measured around 0.2--0.3\,MK falling from 0.17 at solar minimum to 0.11 at
solar maximum. The effect is amplified at higher temperatures (0.3--0.6\,MK) with a stronger anti-correlation
and the ratio falling from 0.16 at solar minimum to 0.08 at solar maximum.
The values we find at solar minimum are too low to solve the solar modeling problem.
\end{abstract}


%% ------------------------------------------------------------------------------------------
%% --- BODY ---------------------------------------------------------------------------------
%% ------------------------------------------------------------------------------------------
 
\section{introduction}

\begin{figure*}
  \centerline{\includegraphics[width=0.95\linewidth]{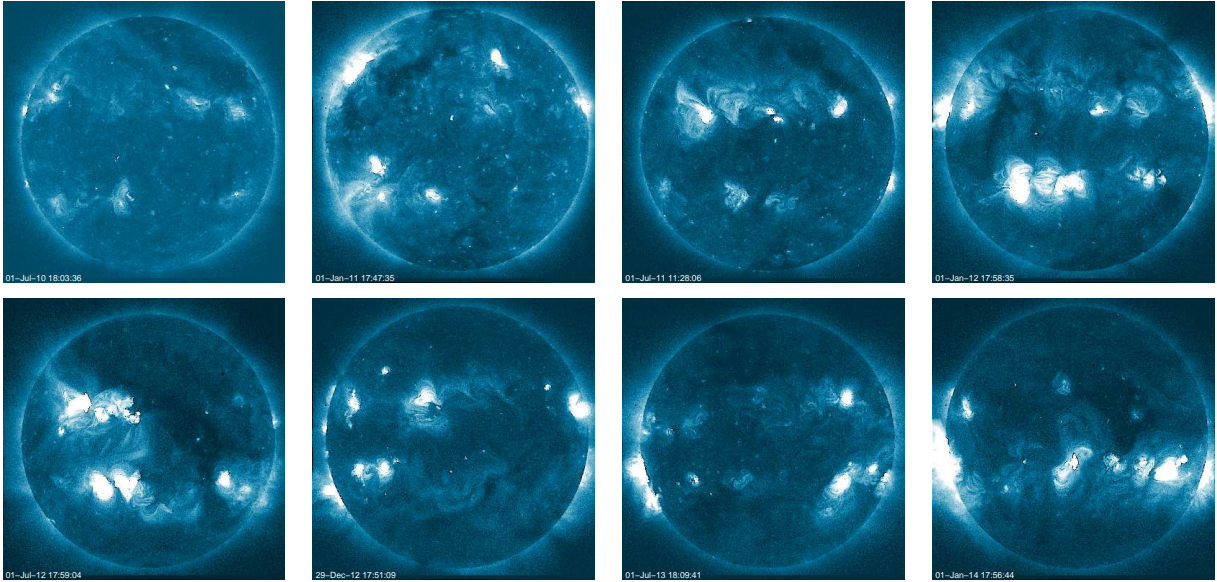}}
  \vspace{-0.1in}
\caption{ A sample of XRT Al-Poly images taken at 6-month intervals covering the period analyzed in
this work between July 2010 and January 2014.
\label{fig:fig1}}
\end{figure*}

The chemical composition of the Sun is a fundamental benchmark for many branches of astrophysics
from solar system evolution to the heating of stellar coronae. An accurate knowledge of solar
elemental abundances is a prerequisite for many investigations, but our understanding
of the solar composition has been called into question by our inability to reconcile  
helioseismological observations of the Sun's interior with the predictions of theoretical models.
This issue has become known as the $``$solar abundance problem$"$, and refers to the discrepancy
introduced by upgraded calculations which lowered the photospheric abundances of elements such 
as carbon, nitrogen, oxygen, and neon \citep{asplund_etal2004}. These elements are significant contributors to the opacity of the solar
interior, so the reduced abundances also reduced the opacity of the radiative interior, and thus
created a mismatch, for example, between the position of the base of the convective zone compared
to that derived from helioseismic measurements \citep{bahcall_etal2005a}. Previously the observations and models were in excellent agreement
not only in the depth of the convection zone but also, for example, the sound speed profiles \citep{bahcall_etal1997}.

There can be little doubt that the revised abundances based on 3-D numerical hydrodynamic models that 
relax assumptions such as local thermodynamic equilibrium in the spectral line formation, and utilize
improved atomic data, are more realistic. They are able to capture
the topology, spatial, and temporal scales of solar granulation, for example \citep{nordlund_etal2009}, none of which
could be reasonably modeled with the previous 1-D hydrostatic simulations. 
While future improvements can
be expected, such as incorporating 
the influence of photospheric magnetic fields that have only recently
started to be included \citep{fabbian_etal2010,fabbian_etal2012}, attempts to solve the problem have focused
on the abundances that are not inferred directly from the models.

While the photospheric abundances of carbon, nitrogen, and oxygen can be determined by comparison with the absorption lines they
produce in the visible spectrum, there are no detectable neon lines. So the neon abundance is inferred
indirectly by, for example, measuring the Ne/O abundance ratio using UV or X-ray spectra, and converting
the ratio to a neon abundance using the known oxygen abundance. 
This makes the determination of the photospheric
abundance of neon in the Sun much less certain than that of the other elements. 

The measurements are further complicated by the first ionization potential (FIP) effect \citep{pottasch_1963}. The composition
of the solar photosphere and corona are different: elements with a low FIP ($\leq10$\,eV) are enhanced in 
the corona and slow speed solar wind by factors of 2--4, while high-FIP elements retain their photospheric
abundances \citep{meyer_1985,feldman_etal1992}. Since neon (FIP = 21.6 eV) and oxygen (FIP = 13.6 eV) are both usually assumed to be high FIP elements, they
are in principle unaffected. In some circumstances, however,
variations are seen in different features such as quiescent active regions \citep{schmelz_etal1996}, 
and oxygen also apparently sometimes behaves like a low-FIP element with respect to neon in long-lived structures
such as coronal streamers \citep{landi&testa_2015}, or the slow solar wind \citep{shearer_etal2014}. 

Taking these issues into account, \citet{antia&basu_2005} and \citet{bahcall_etal2005b} proposed that if the 
photospheric abundance of neon were 
raised by a factor of 2.5 or more then the solar models would come back into agreement with the seismic observations.
There is, however, no obvious reason to raise the neon abundance. The standard value for neon prior to the
revision came from weighted means of solar wind, solar energetic particle, and flare spectra \citep{grevesse_etal1996}.
Furthermore, the post-revision of \citet{asplund_etal2009} was inferred from EUV 
spectroscopic measurements of the solar transition region between the photosphere and corona by
\citet{young_2005}, 
where no elemental fractionation is observed. Note, however, that very recent measurements by \citet{brooks_etal2017} show evidence of
fractionation even at temperatures associated with the upper transition region ($\sim$0.5\,MK). 

From a sample of X-ray observations of cool stars by {\it Chandra}, however, \citet{drake&testa_2005}
brought forward some evidence in support of this higher value. They
examined a sample of nearby solar-like stars, and found that a value of 0.41 for the Ne/O abundance ratio for the Sun
would be in agreement with the other stars in their sample. This is a factor of 2.3--2.7 larger than the standard Ne/O 
abundance ratio of 0.15--0.18. \citet{asplund_etal2005} argued that the stellar
sample of \citet{drake&testa_2005} was skewed towards more active stars, which show larger inverse FIP effects,
a depletion of high-FIP elements rather than an enhancement of low-FIP elements \citep{drake_etal2001}, and that a 
value around 0.2 would be more reasonable for low activity stars similar to the Sun. Without a reliable solar measurement,
however, it was difficult to settle this issue.

\citet{schmelz_etal2005} re-examined older spectroscopic data from US air force satellites and naval rocket flights in the
1960s and 70s. The measurements of full disk integrated
spectra mimic what the Sun would look like if it were viewed as a star. These $``$Sun-as-a-star$"$ observations
allowed a direct comparison with the stellar measurements, but were found to be consistent with the standard
Ne/O abundance ratio of around 0.15, apparently closing this particular avenue of exploration.

Recently, a new pathway to reconciliation has potentially emerged. \citet{shearer_etal2014}, found that the Ne/O abundance ratio varies
with the solar cycle in slow solar wind streams ($<$ 400\,km s$^{-1}$), and a similar cyclic variation
was subsequently observed by \citet{landi&testa_2015} in coronal streamers. Although the magnitude of the variation in the slow
wind is only around 40\%, the variation in streamers is a factor of 2.5. Furthermore, based on the idea that the FIP effect
was suppressed during the unusually 
extended solar minimum from 2007--2010, \citet{landi&testa_2015} argue that the higher value of 0.25 for the Ne/O abundance ratio
measured at that time is closer to the true photospheric abundance. In fact, more recent 
studies of the abundances of oxygen and other light elements and comparison to helioseismic and solar neutrino data by
\citet{antia&basu_2011} and \citet{Villante_etal2014} suggest that the magnitude of the variation seen by \citet{landi&testa_2015}
may be enough to restore agreement with the seismic data. 

This conjecture remains unproven, and there are other dissenting arguments 
from \citet{grevesse_etal2013}, who point out that some recent measurements of the oxygen photospheric abundance
do not completely account for subtle blending spectral lines \citep{caffau_etal2011}, and, in fact, that when these blends are taken 
into account, the
O abundance is reduced. Conversely, a very recent re-analysis of quiet Sun observations using new atomic data supports a high
value of 0.24 for the Ne/O abundance ratio \citep{young_2018}.

\begin{deluxetable}{lclc}
\tabletypesize{\scriptsize}
\tablewidth{0pt}
\tablecaption{Line Selection}
\tablehead{
\multicolumn{1}{l}{Line ID} &
\multicolumn{1}{c}{$\log$ T$_{p}$} &
\multicolumn{1}{c}{Line ID} &
\multicolumn{1}{c}{$\log$ T$_{p}$} \
}
\startdata
  O III 599.598\AA &  5.02 &  Si VII 275.352\AA &  5.80 \\
   O IV 554.510\AA &  5.22 &   Fe IX 171.073\AA &  5.92 \\
    O V 629.730\AA &  5.38 &    Fe X 174.532\AA &  6.06 \\
   Ne V 572.311\AA &  5.44 &    Fe X 177.239\AA &  6.06 \\
   Ne V 569.820\AA &  5.44 &    Si X 258.375\AA &  6.14 \\
   O VI 1037.64\AA &  5.48 &   Fe XI 180.401\AA &  6.14 \\
   O VI 1031.93\AA &  5.48 &    Si X 261.040\AA &  6.14 \\
 Ne VII 465.221\AA &  5.72 &  Fe XII 195.119\AA &  6.20 \\
Fe VIII 131.240\AA &  5.76 & Fe XIII 202.044\AA &  6.24 \\
 Mg VII 278.402\AA &  5.80 &  Fe XIV 211.316\AA &  6.30 \\
Ne VIII 780.380\AA &  5.80 &   Fe XV 284.160\AA &  6.34 \\
Ne VIII 770.420\AA &  5.80 &  Fe XVI 262.984\AA &  6.44
\enddata
\tablenotetext{}{T$_p$ - peak of the G(T$_e$) function.}
\label{table1}
\end{deluxetable}

Recently, \citet{brooks_etal2017} detected a solar cyclic variation in the magnitude of the FIP effect in full disk
integrated Sun-as-a-star 
spectra obtained by the Solar Dynamics Observatory. Motivated by this finding, the detections of solar cyclic variations in
the Ne/O abundance ratio in
the slow wind and detailed atmospheric structures, and the continuing controversy of the solar abundance problem, we decided
to investigate whether the Ne/O abundance ratio depends on the solar cycle in Sun-as-a-star data. Our analysis expands on the 
work of \citet{schmelz_etal2005}, with much more extensive data coverage of the solar cycle, and more recent
and updated atomic data. The detection of a cyclic variation of the FIP effect in the Sun-as-a-star data suggests this may
be a promising route for solving the solar abundance problem. It also allows a direct comparison with the stellar observations, 
and could potentially reconcile the Sun-as-a-star Ne/O abundance ratio with that measured in nearby stars.

\section{Methods}

We use SDO/EVE \citep[Extreme ultraviolet Variability Experiment,][]{woods_etal2012} data for our analysis. EVE includes
several instruments for measuring the solar EUV irradiance, including twin
grating near-normal incidence spectrographs that observe in two wavelength bands from 50--370\,\AA\, and 350--1050\,\AA\, with a coarse spectral
resolution of 1\,\AA. The short wavelength band is recorded by MEGS-A (Multiple EUV Grating Spectrographs) and the long
wavelength band by MEGS-B. The observations are full disk integrated spectra. We use MEGS-A and MEGS-B
level-2 v.5 spectra in this work. These data were downloaded from the Science Processing Operations Center of 
the Laboratory for Atmospheric and Space Physics in Colorado. The data products are fully calibrated
spectra merged from the two wavelength bands.
They are 10 second integrated spectral irradiance measurements that are provided
as hourly datasets. We computed daily spectra by median filtering all of the spectra in the hourly 
datasets within each 24 hour period as in \citet{warren_2014}. We also converted the irradiance measurements from the SI units W m$^{-2}$ nm$^{-1}$ to 
intensities in cgs units of erg
cm$^{-2}$ s$^{-1}$ sr$^{-1}$ by accounting for the solid angle, $\pi r^2 / R^2$, where $r$ is the solar radius and $R$ is the Earth-Sun distance.
The observations cover the period from April 2010 to May 2014 (1475 days in total).
We show context images covering this period from the {\it Hinode} X-ray Telescope \citep[XRT,][]{golub_etal2007} 
in Figure \ref{fig:fig1}.

EVE observes a large number of spectral lines covering a wide range in temperatures and we selected 
a set of lines that are relatively blend free (Table \ref{table1}). 
The selection includes lines of \ion{O}{3}--\ion{O}{6}, \ion{Ne}{5}--\ion{Ne}{8},
\ion{Si}{7}, \ion{Mg}{7}, \ion{Si}{10}, and \ion{Fe}{9}--\ion{Fe}{16}, which cover a temperature
range from 0.1\,MK--2.75\,MK.
We extracted the intensities for all the spectral lines shown in Table \ref{table1} for
each of the daily datasets. 
To achieve this, we integrated the intensities between pre-defined wavelength
limits chosen to minimize the impact of surrounding blends. 
We assessed the likely impact of blends in our previous work \citep{brooks_etal2017} using much higher
spectral resolution observations of quiet and active regions from the EUV Imaging Spectrometer on the 
Hinode satellite \citep{brown_etal2008}. Blending lines within 1.5\,\AA\, of the line positions mostly
produce a contribution that is comparable to the assumed uncertainties. See \citet{brooks_etal2017}
for details and exceptions. 

We assume a 20\% uncertainty in the intensities in our analysis \citep{woods_etal2012}.
This is probably conservative for the lines below 750\,\AA\, where the calibration uncertainty is likely smaller,
but takes some account of uncertainties at longer wavelengths on MEGS-B that are assumed to be larger.
For example, by cross-calibration
between MEGS-B and the Solar EUV Experiment (SEE) on the TIMED \citep[Thermosphere Ionosphere Mesosphere Energetics and Dynamics,][]{woods_etal2005}
mission, \citet{milligan_etal2012} were able to derive a correction factor for the 750--921\,\AA\, wavelength range
during 2011 February. Applying their correction to our daily spectra taken, for example, on February 15, results in an average
decrease of $\sim$20\% for the extracted intensities of the \ion{Ne}{8} spectral lines we use in that wavelength range. This is comparable to the 
uncertainty we assume.

Having obtained the intensities, we followed the method of \citet{landi&testa_2015} to compute the Ne/O abundance ratio. The ratio of the
intensities of two spectral lines, one each emitted by neon and oxygen, is given by
\begin{equation}
{I_{Ne} \over I_{O}} = {\int G_{Ne} (T,n) \phi (T) dT  \over \int G_{O} (T,n) \phi (T) dT},
\end{equation}
where T is the electron temperature, n is the electron density, $\phi$ is the distribution of coronal plasma with 
temperature, or differential emission measure (DEM),
and $G(T,n)$ is the line emission contribution function. $G(T,n)$ contains all of the necessary atomic parameters and
the abundance of the element, $A(Z)$, is separable from it as
\begin{equation}
G(T,n) = A(Z) g(T,n). 
\end{equation}

Given the electron density and emission measure distribution, the ratio $I_{Ne}/I_{O}$ can be calculated as a function of
the Ne/O abundance ratio: $A(Ne)/A(O)$. The initial abundance ratio is assumed, and comparison with the
observed ratio produces a correction to the assumed ratio. We use two sets of neon and oxygen spectral lines 
i.e. \ion{Ne}{5} 572.311\,\AA,
\ion{Ne}{5} 569.820\,\AA, and
\ion{O}{4} 554.510\,\AA\, covering the 0.17--0.28\,MK\, temperature range, and 
\ion{Ne}{8} 770.420\,\AA,
\ion{Ne}{8} 780.380\,\AA,
\ion{O}{6} 1031.93\,\AA, and
\ion{O}{6} 1037.64\,\AA\, covering the 0.3--0.63\,MK\, temperature range. The second set of lines are 
the same lines as in \citet{landi&testa_2015}.

To calculate the electron density and emission measure (EM) distribution, we follow a modified version of the
methodology we used in previous work \citep{brooks&warren_2011,warren_etal2011,brooks_etal2015}.
We compute the electron density using the \ion{Si}{10} 258.375/261.04 diagnostic ratio. We then use this value
to compute the $G(T,n)$ functions. Together with the observed
intensities, the $G(T,n)$ functions are used to calculate the EM distribution for every daily spectrum. 
All the spectral lines in Table \ref{table1} were used for the EM calculations except \ion{Ne}{8} 770.420\,\AA\,
and \ion{Ne}{8} 780.380\,\AA\, (see section \ref{s32}).
Technically, we compute the EM using the
Monte Carlo Markov Chain (MCMC) algorithm available in the PINTofALE software package \citep{kashyap&drake_1998,kashyap&drake_2000}.
This technique finds the best-fit solution from 100 simulations where the observed intensities are randomly perturbed within the 
uncertainty.
The $G(T,n)$ functions are calculated using the CHIANTI database v.8 \citep{dere_etal1997,delzanna_etal2015}.
We adopt the coronal abundances of \citet{feldman_etal1992} to determine the most accurate EM, 
though the method
is insensitive to the assumed abundances since we derive a correction to them. 

Once we have the EM for each dataset, we compute the correction to the assumed abundances using
the four independent $I_{Ne}/I_{O}$ line ratios. 
We take the average of the correction factors calculated for each of the four ratios. 

We also compute the standard deviation of a distribution of 1000 random trials where each one 
finds the best fit
EM solution to 100 Monte Carlo simulations, as above, and calculates a value for the $A(Ne)/A(O)$ ratio. 
We take this standard deviation as a measure of the uncertainty in the measurements. 

\section{results and discussion}

\begin{figure}
  \centerline{\includegraphics[width=0.95\linewidth]{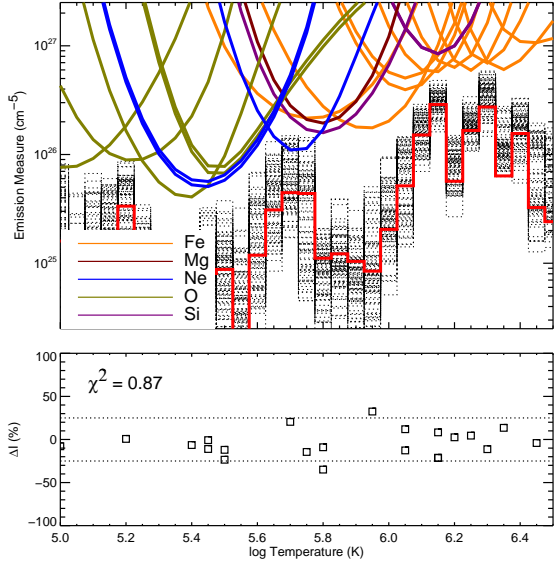}}
  \vspace{-0.1in}
\caption{EM solution for the daily spectrum taken on 4th May 2012. The best-fit EM
solution is shown in red, and the Monte Carlo simulations are shown by the grey dotted lines. The
colored lines are emission measure loci curves for all the spectral lines used in the calculation.
The lower panel shows the differences between the
observed and EM calculated intensities as a percentage of the observed intensity. They are plotted
at the formation temperature of the relevant line. We show the full line-list
with formation temperatures for cross-checking in Table \ref{table1}. The reduced $\chi^2$ value
for the solution is shown in the legend.
\label{fig:fig2}}
\end{figure}

\begin{figure}
  \centerline{\includegraphics[width=0.95\linewidth]{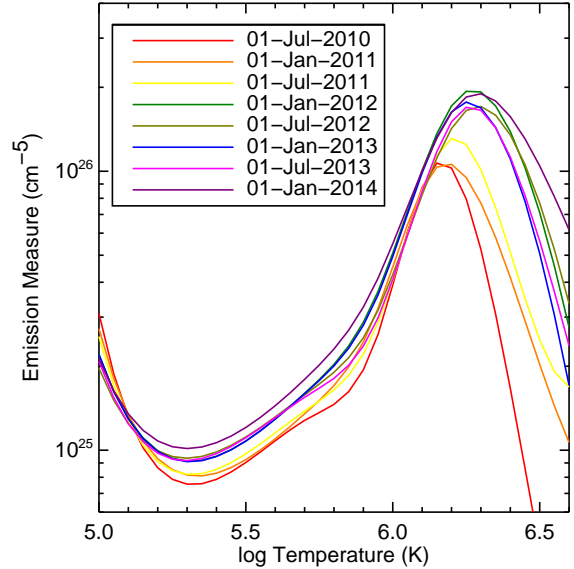}}
  \vspace{-0.1in}
\caption{ A sample of smoothed EM solutions taken at 6-month intervals showing the effect of the rise in
solar activity between July 2010 and January 2014. The curves are rainbow color coded from red
near solar minimum to violet near solar maximum (c.f. Figure \ref{fig:fig1}).
\label{fig:fig3}}
\end{figure}

\begin{figure*}
  \centerline{\includegraphics[width=0.50\linewidth]{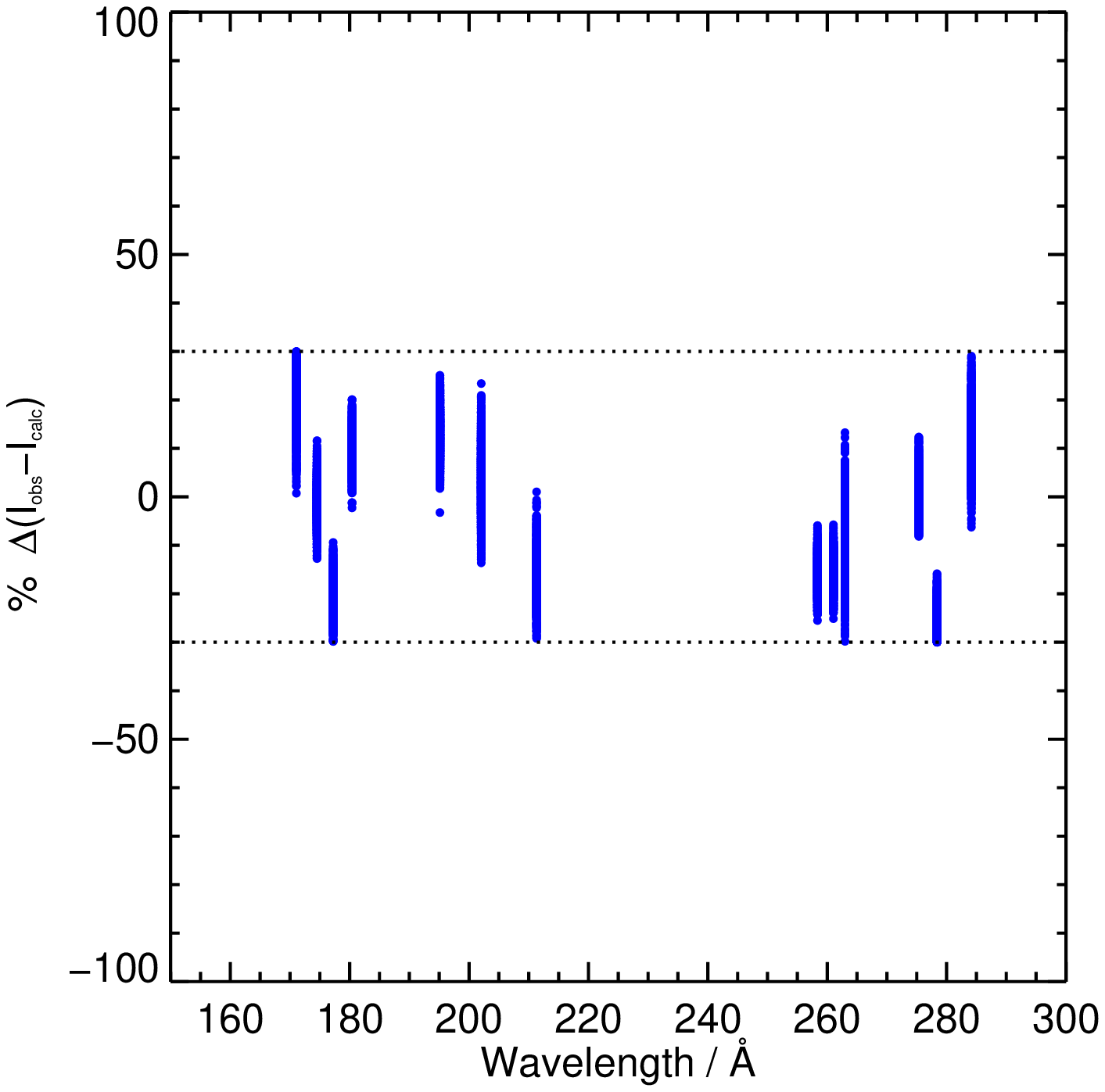}
              \includegraphics[width=0.50\linewidth]{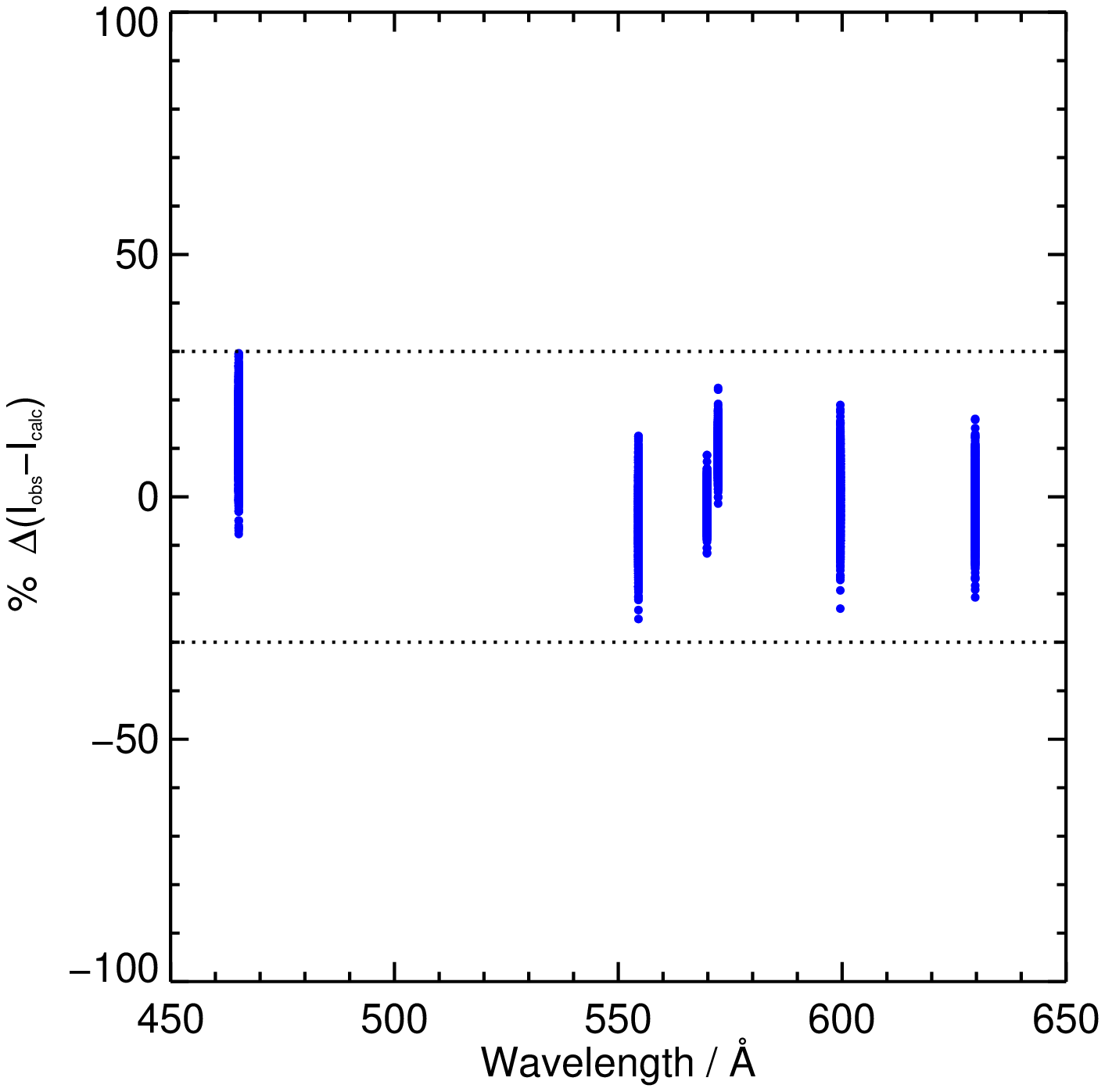}}
  \vspace{-0.1in}
\caption{The difference between observed and calculated intensities expressed as a percentage of the
observed intensity for all the spectral lines used in the emission measure analysis, except the those beyond 1000\,\AA.
The data are plotted at the wavelength
of the relevant line. We show the full line-list with wavelengths for cross-checking in Table \ref{table1}.
Results are shown for all daily spectra analyzed (1103 observations).
\label{fig:fig4}}
\end{figure*}

\subsection{ Emission Measure distributions}

In Figure \ref{fig:fig2} we show an example of the EM distribution computed from the extracted
intensities for the spectral lines shown in Table \ref{table1}. The EM loci curves in the Figure
show the upper constraints on the EM. They were calculated assuming all of the intensity is 
emitted from the peak formation temperature i.e. by dividing the intensity by the contribution function.
The reduced $\chi^2$ value is close
to one, which indicates that the EM model is well fitting the observed intensities. The differences between 
the EM calculated and observed intensities are less than 25\% for nearly all the lines (lower panel).

We performed a similar calculation for every daily spectrum in the complete dataset. There is considerable structure
in the MCMC calculations, and some differences depending on which lines are included in the analysis (see below). So for clarity of display, 
and ease of comprehension of the global trends, we show a selection 
of EM distributions at six month intervals from our sample, that have been post-processed by spline interpolation so that they are smoother. That is, 
the MCMC EM calculation for the date was used as input to a spline interpolation algorithm where several temperature/EM knot points
are interactively selected, and the solution is calculated by minimizing the differences between the observed and predicted intensities
\citep{warren_2005}. The displayed distributions show the evolution of the EM from near the minimum of solar
cycle 23 in July, 2010, to beyond the maximum of solar cycle 24 in January, 2014. The XRT images
in Figure \ref{fig:fig1} were taken as close in time as possible to the dates shown in Figure \ref{fig:fig3}.
There is an enhancement
in EM at all temperatures as the cycle activity increases, with the peak EM increasing from $\log$ EM = 26.03
to 26.28 (a factor of nearly two). 
The coronal peak temperature also evolves from $\log$ T = 6.15 to 6.30,
corresponding to an increase from 1.4\,MK near solar minimum to 2\,MK near activity maximum.
The EM shows much larger increases at higher temperatures, but the calculations are most uncertain in this
region due to a lack of high temperature constraints above $\log$ T = 6.5 in the EVE spectra. 

These apparent trends are consistent with the independent analysis of {\it SDO} data by \citet{morgan&taroyan_2017},
who also note a similar increase in coronal mean temperature from 1.4\,MK to 1.8\,MK, primarily driven by
the presence and increase in a high temperature component near $\log$ T = 6.5 (3\,MK). \citet{morgan&taroyan_2017}
suggest that this high temperature component also leads to a solar cyclic variation in the EM distribution
below $\log$ T = 6.2 (1.6\,MK), however, a separate analysis of EVE spectra reports that there is no variation
with the solar cycle at these temperatures \citep{schonfeld_etal2017}. Our results are consistent with \citet{schonfeld_etal2017}:
as Figure \ref{fig:fig3} shows, the EM distribution is very similar up to at least $\log$ T = 6.15 at all times.
From previous work examining different spatial regions, we also do not expect much variation at these temperatures.
The quiet Sun coronal EM distribution is very similar regardless of the observing instrument when averaged over
significant areas \citep{brooks&warren_2006,brooks_etal2009}. Indeed \citet{brooks_etal2009} suggest that it has
a universal character, driven by the radiating and conducting properties of the plasma. In this picture, the
explanation for the different distribution near solar maximum found by \citet{morgan&taroyan_2017}, is that the 
observations at this time are not of truly quiet Sun. The EM distribution averaged over regions of quiet Sun peaks at 1.1\,MK and has 
fallen by two orders of magnitude already by 2\,MK; see for example Figure 6 of \citet{brooks_etal2009}. So the composite quiet Sun EM distribution 
averaged over most of the disk is contaminated by 
higher temperature (3\,MK) activity near solar maximum, which may well be due to decaying active regions as \citet{morgan&taroyan_2017}
suggest. The EM distribution near solar maximum should still have a universal shape, but it will be different
from the usual quiet Sun EM shape because the high temperature (different density) component is accessing a 
different part of the radiative loss function, which is not seen near solar minimum. 

These additional
data from other instruments certainly help to definitively pin down the changes at these temperatures, but
the abundance diagnostics we use are sensitive to cooler plasma where the EM distribution is well constrained.

To ensure that our results are robust for all the daily spectra, we examined the differences between the 
observed and calculated intensities for all the spectral lines in the complete dataset. Figure \ref{fig:fig4}
shows these differences calculated as a percentage of the observed intensities for all the daily spectra
we used. This plot is analogous to the lower panel of Figure \ref{fig:fig2}. If these differences were 
greater than 30\% for any of the lines in a particular daily spectrum, then that date was excluded 
from our subsequent analysis. 

\subsection{ Comment on the accuracy of the Li-like spectral lines }
\label{s32}

There is a caveat to our threshold condition. While the atomic structure of Li-like ions is very simple (three
electrons only) and lines emitted from these ions are therefore expected to be spectroscopically accurate, they are well
known to sometimes show anomalously large intensities \citep{dupree_1972,judge_etal1995}. Based on off-limb observations 
from Hinode/EIS, where the solar corona is close to isothermal and so atomic data discrepancies are readily
apparent, \citet{warren_etal2016} found that the contribution functions for some \ion{O}{6} lines need to be 
adjusted upward by a factor of 3.4 to bring them into agreement with other lines. Other work has found that
these discrepancies are wavelength dependent \citep{muglach_etal2010}, and several studies have successfully
used the longer wavelength \ion{O}{6} lines we use here for composition studies \citep{feldman_etal1998,landi&testa_2015}. 

In our analysis, we also found that the Li-like \ion{O}{6} 1031.93\,\AA, \ion{O}{6} 1037.64\,\AA, \ion{Ne}{8} 770.42\,\AA, and
\ion{Ne}{8} 780.38\,\AA\, observed line intensities were several factors brighter than the intensities calculated from the 
EM distribution. 
Excluding the \ion{Ne}{8} lines and adjusting the
$G(T,n)$ functions for \ion{O}{6} 1031.93\,\AA\,
and \ion{O}{6} 1037.64\,\AA\, 
upward by the factor of 3.4 determined from EIS off-limb spectra by \citet{warren_etal2016} improves the minimization of the EM solution, and helps to anchor
the distribution around their formation temperatures.
We therefore make the adjustment solely for the purpose of determining the best EM solution.
We make no adjustment for the subsequent abundance ratio correction since \ion{Ne}{8} is also Li-like and it is necessary
to treat them consistently.

Note also that the \citet{warren_etal2016} adjustment factor was computed using off-limb
EIS spectra taken near solar minimum in 2007, November. Our analysis suggests that this factor is also representative of the rest
of the solar cycle, since the computed \ion{O}{6} line intensities are maintained within 25\% of the observed intensities
in all the daily spectra. If there was a cycle dependent variation greater than 25\% in this adjustment factor, perhaps due to the 
presence or absence of the high temperature component in the emission measure found by \citet{morgan&taroyan_2017},
then it would be detected in our analysis as a discrepancy between the calculated and observed intensities.

\begin{figure}
  \centerline{\includegraphics[width=0.95\linewidth]{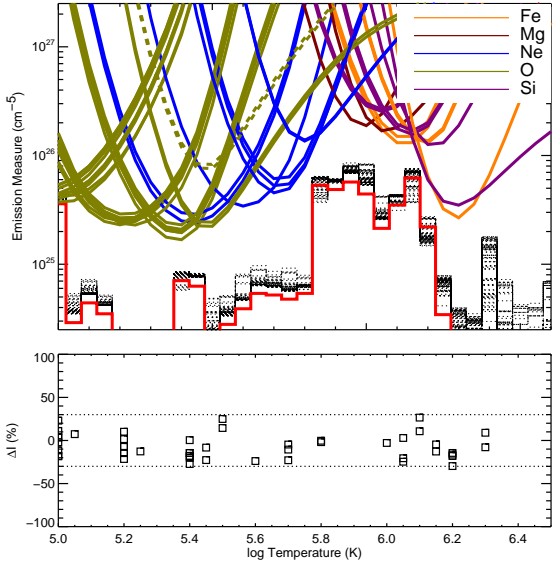}}
  \vspace{-0.1in}
\caption{EM solution for the quiet Sun spectrum from combined CDS \& SUMER data. The description of
\ref{fig:fig2} also corresponds to this Figure and the curves follow the same conventions except
for the dashed curve. This is the EM loci curve for the Li-like \ion{O}{6} lines before 
adjustment (see text).
\label{fig:fig5}}
\end{figure}

It is interesting that the same adjustment factor works well for both our EVE analysis and the independent EIS off-limb spectral analysis. 
For a further cross-check of consistency we have re-examined data of the quiet Sun analyzed by \citet{warren_2005}. \citet{warren_2005}
combined spectra obtained by the SOHO Coronal Diagnostic Spectrometer \citep[CDS,][]{harrison_etal1995} and Solar Ultraviolet Measurements of
Emitted Radiation \citep[SUMER,][]{wilhelm_etal1995} instrument. Twenty spectral atlas observations of the quiet Sun were processed and
added together to form the CDS part of a composite spectrum covering the 308--381\,\AA\, and 513--633\,\AA\, wavelength ranges. For SUMER,
central meridian scans covering the 660--1500\,\AA\, wavelength range were used. The data are much higher spectral resolution than our EVE
spectra. The CDS spectral resolution is 80\,m\AA\, in the shorter wavelength band, and 140\,m\AA\, in the longer wavelength band. The SUMER 
spectral resolution is about 45\,m\AA.

\citet{warren_2005} provided the line intensities for the composite spectrum in Tables 1 \& 2. We selected 48 emission lines from their tables, and
computed the emission measure distribution
from these intensities using the same methodology as we used for the EVE spectra. Our selection criteria was to cover the same temperature 
range as the EVE spectra, and exclude lower temperature lines that are potentially affected by optical depth effects \citep[see eg. the 
analysis of ][]{brooks_etal2000}. We also excluded Li-like and Na-like lines, with the exception of the \ion{O}{6} lines we were testing. This was so that 
any discrepancy in the intensities for the \ion{O}{6} lines would be unaffected by the anomalous behavior of other species. Finally, we 
excluded lines that were not reproduced by the emission measure calculation. We therefore used lines from \ion{O}{3}--\ion{O}{6}, 
\ion{Ne}{5}--\ion{Ne}{8}, \ion{Si}{9}--\ion{Si}{12},
\ion{Fe}{11}--\ion{Fe}{14}, and \ion{Mg}{10}.

We show the resulting EM distribution in Figure \ref{fig:fig5}. The lower panel shows that all the 48 emission line intensities are reproduced, including
the \ion{O}{6} 1031.94\,\AA\, and \ion{O}{6} 1037.64\,\AA\, intensities. The Figure shows the results including the factor 3.4 adjustment. We show the unadjusted
EM loci curves for the \ion{O}{6} lines with the dashed olive curve. Without the adjustment, these lines are clearly discrepant. These results from the
CDS \& SUMER data are consistent with both the off-limb EIS spectra, and our analysis of all the EVE spectra. The same 3.4 adjustment factor works for
all the instruments and data, and this strongly suggests that the problem lies in the atomic physics. The cross-check rules out uncertainties
in the EVE calibration at these wavelengths as the source of the discrepancy.

\subsection{Neon/Oxygen Abundances}

Our final sample includes 1103/1475 daily spectra, and all of the line intensities in Table \ref{table1} and Figure \ref{fig:fig2} are within the 
30\% limit in all of these observations. This is a much more significant dataset of Sun-as-a-star observations sampling
the solar cycle than any previous work. The satellite and 
rocket flight data used in earlier Sun-as-a-star studies had only sparse coverage of the solar cycle \citep{schmelz_etal2005}. 
While the flights in 1963--1975 did cover the minimum and maximum of solar cycle 20, 
and there is some suggestion that the measured $A(Ne)/A(O)$ ratios ranging from 0.18--0.25 correlate at least with the rise of the solar cycle,
the value of 0.20 measured in March 1969 makes it unclear whether there is any cyclic trend. 

\begin{figure*}
  \centerline{\includegraphics[width=0.50\linewidth]{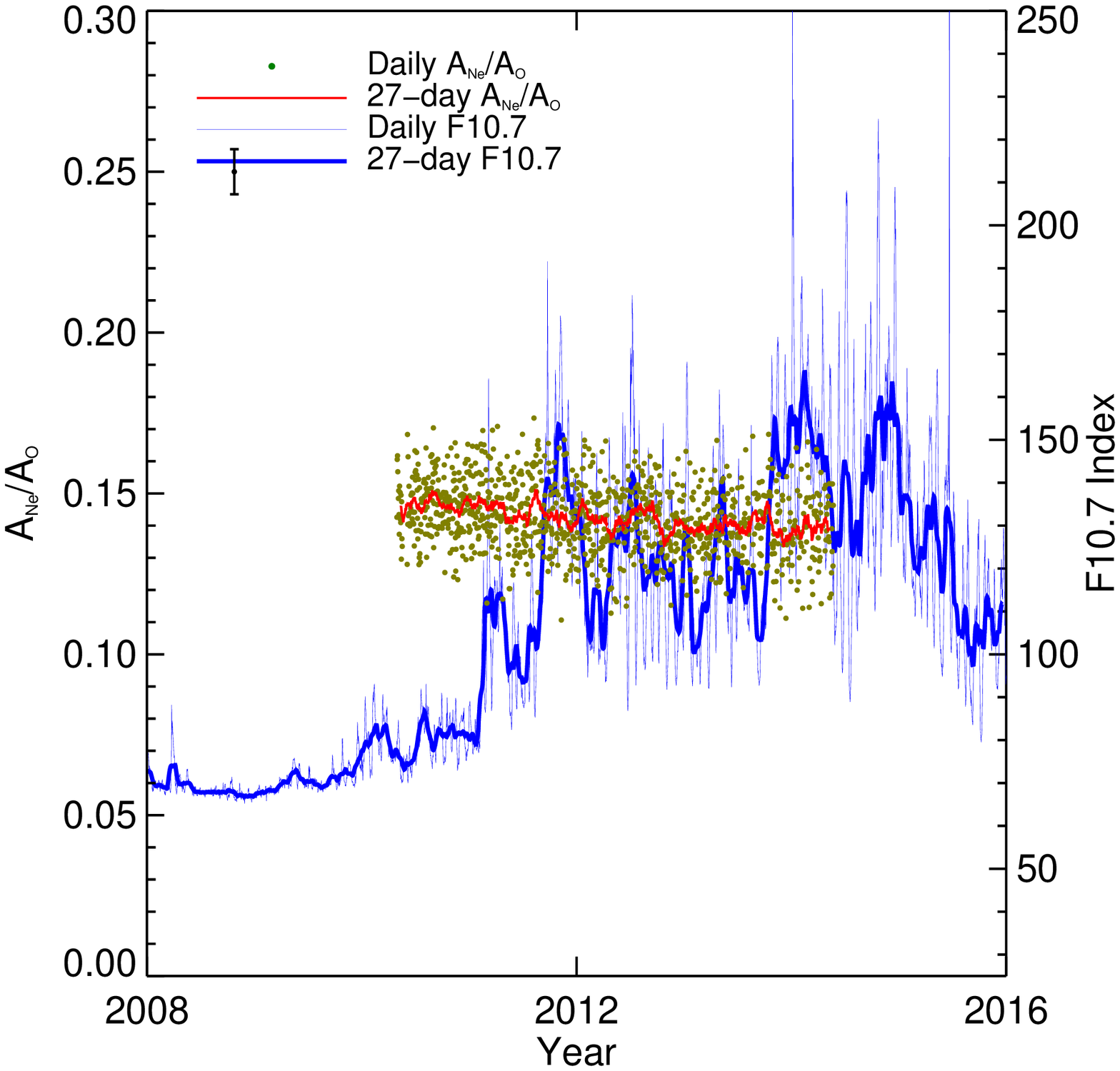}
              \includegraphics[width=0.50\linewidth]{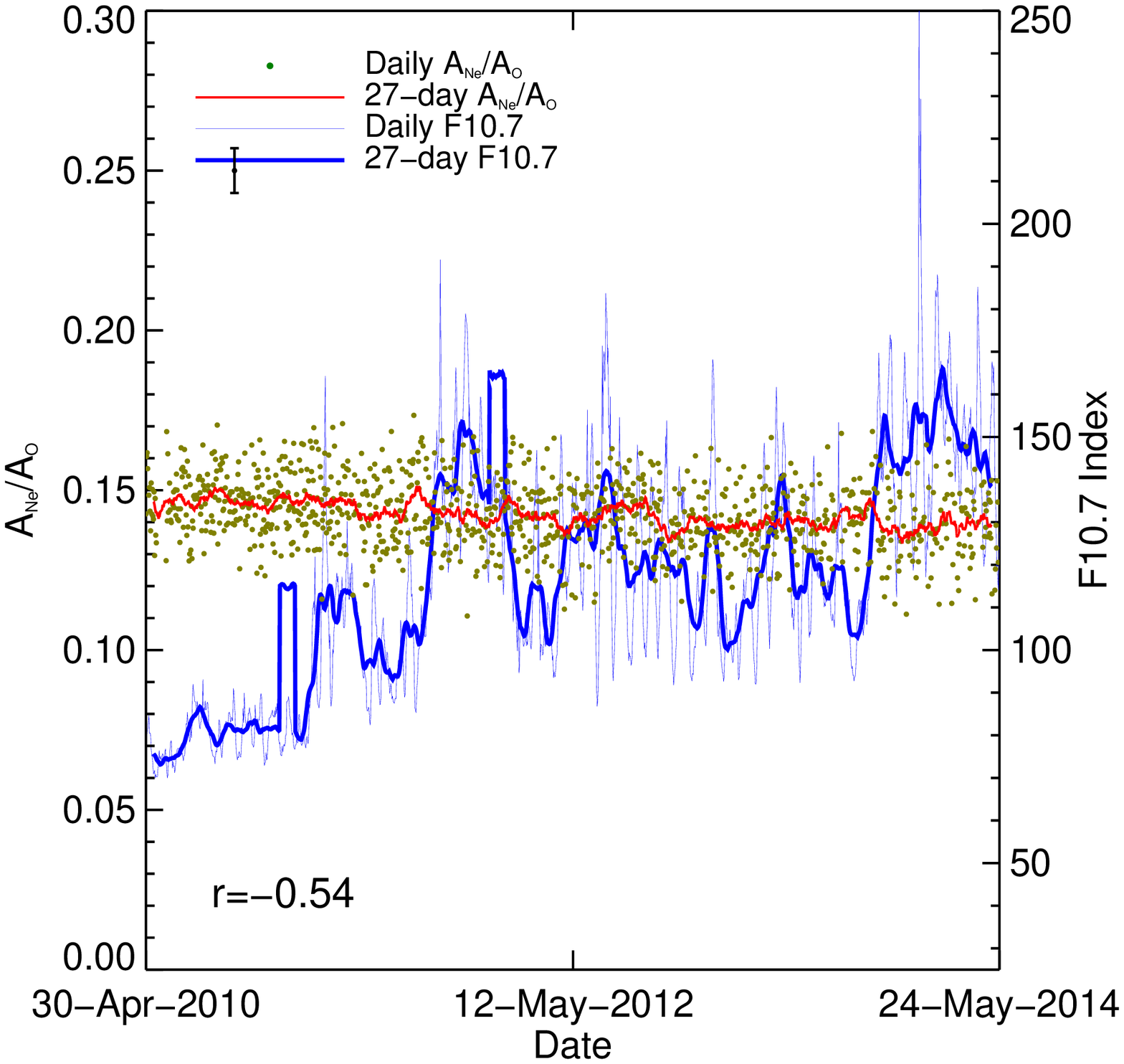}}
  \vspace{-0.1in}
\caption{{\it Left panel}: F10.7\, cm solar radio flux for 2008--2016 (thin blue line) with the Ne/O abundance ratios 
calculated from the \ion{Ne}{5}/\ion{O}{4} spectral lines
overplotted on the dates of the EVE daily spectra (green dots). We also show the 27-day Carrington running
averaged data with the thick blue solid line (F10.7\, cm) and red solid line (Ne/O abundance ratios). 
We show the uncertainty in the measurements with the vertical bar.
{\it Right panel}: Same as the left panel but zoomed in to the period of the EVE data (2010-2014). We give the 
linear Pearson correlation coefficient in the legend. 
\label{fig:fig6}}
\end{figure*}

\begin{figure*}
  \centerline{\includegraphics[width=0.50\linewidth]{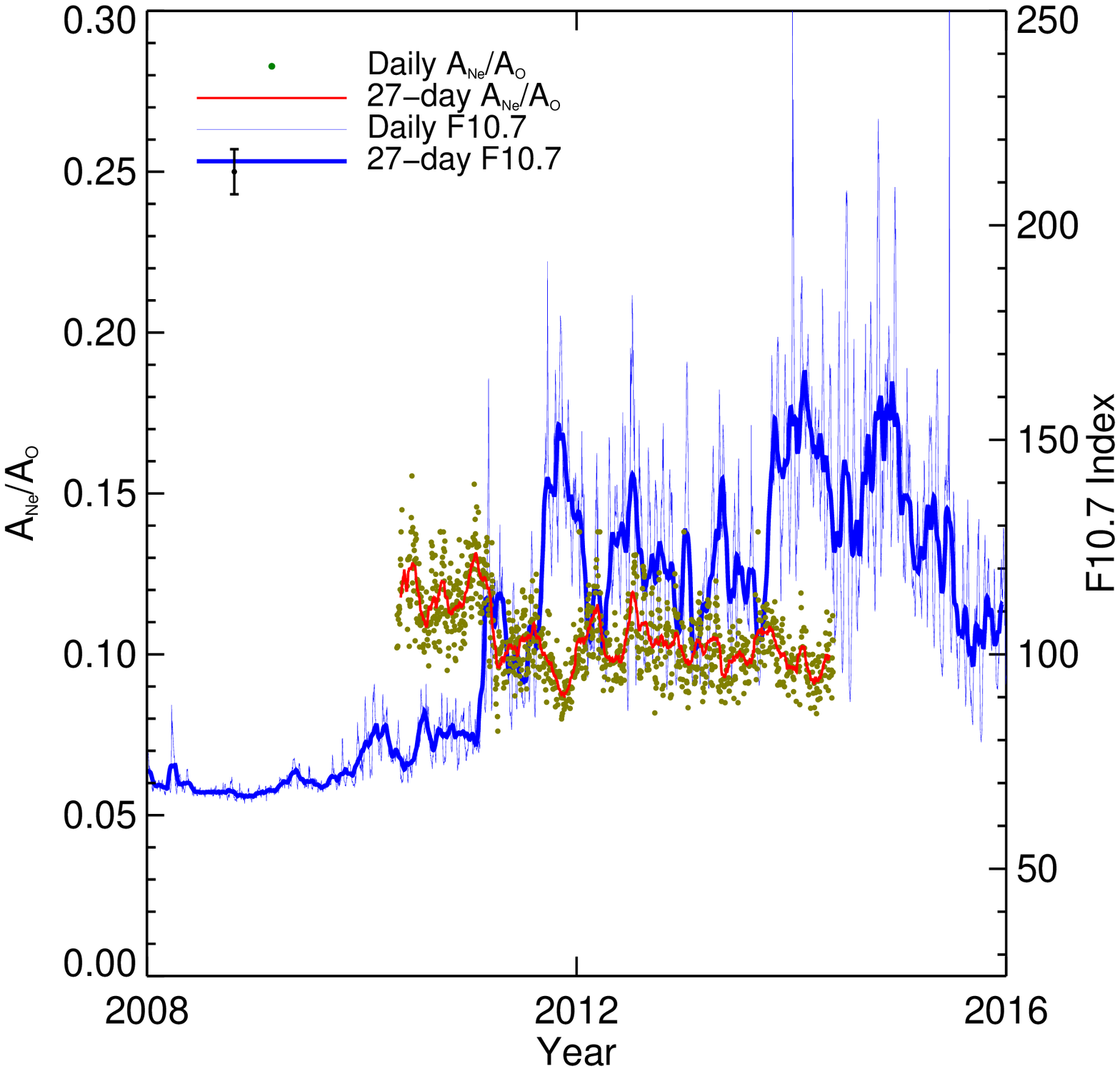}
              \includegraphics[width=0.50\linewidth]{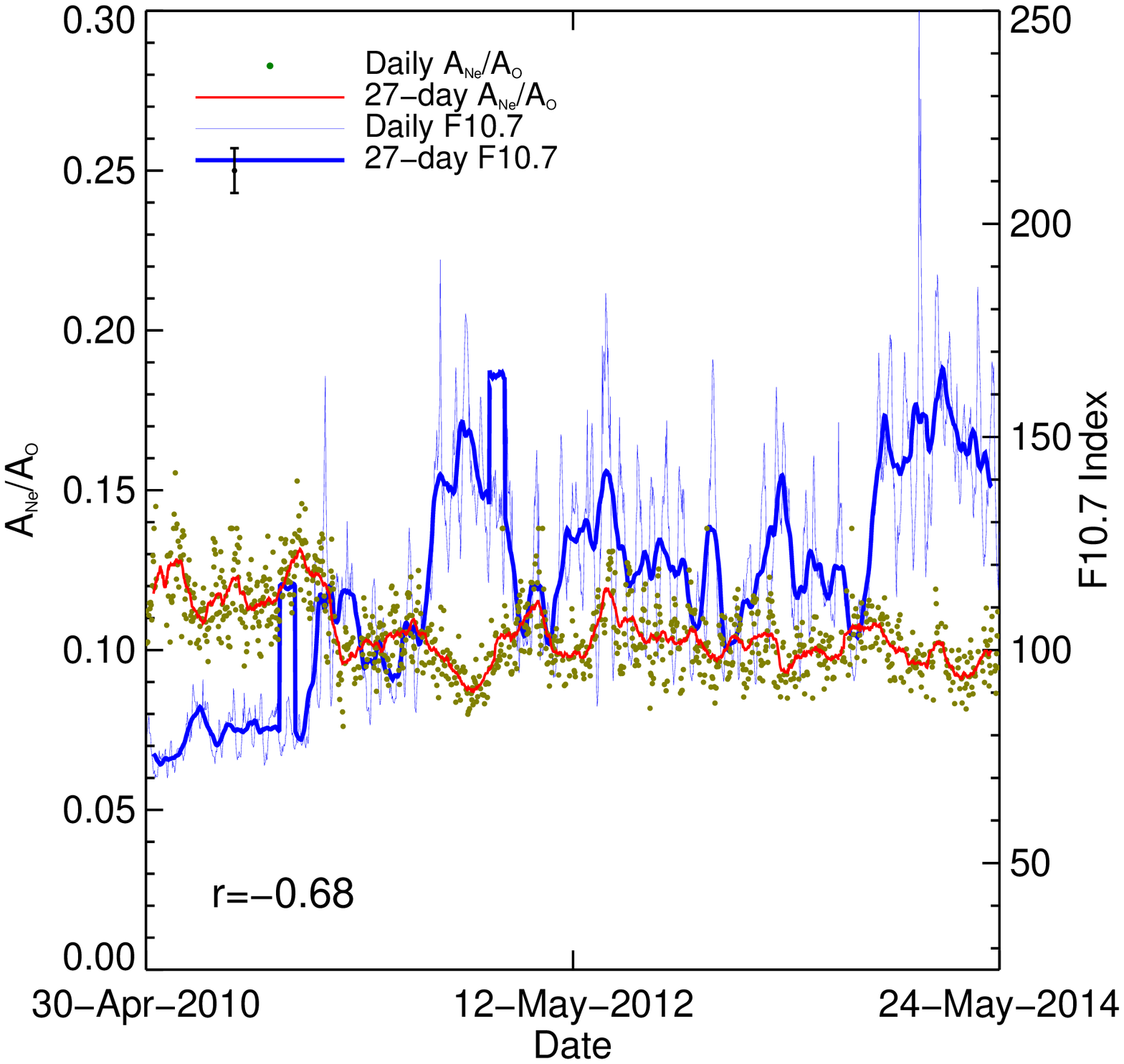}}
  \vspace{-0.1in}
\caption{Same as Figure \ref{fig:fig6} but displaying the results computed from the \ion{Ne}{8}/\ion{O}{6} ratio. 
\label{fig:fig7}}
\end{figure*}

We show our results in Figures \ref{fig:fig6} and \ref{fig:fig7}. 
Figure \ref{fig:fig6} shows the results calculated from the \ion{Ne}{5}/\ion{O}{4} ratio.
The left panels show the 
daily and 27-day Carrington running average F10.7\,cm solar radio fluxes for the period 2008--2016
with our daily-measured $A(Ne)/A(O)$ ratios and 27-day Carrington running average values overlaid. 
The daily $A(Ne)/A(O)$ values fall in the range
0.11--0.17$\pm$0.01, decreasing from the higher value in 2010 near solar minimum when the coronal temperature was measured to be 1.4\,MK, 
to the lower value near solar maximum in 2014 when
the temperature was closer to 2\,MK. The 27-day running average show this fall less clearly, falling slightly from 0.15 in 2010 to 0.13 in 2014. That is,
we find a moderate anti-correlation with the solar cycle. This is perhaps clearer to see in the 27-day running averaged data, and in
the right panel of Figure \ref{fig:fig6} where we have zoomed in to the period of the EVE data.
The strength of the anti-correlation can be 
computed from the linear Pearson correlation coefficient and is -0.54 for these data.

Figure \ref{fig:fig7} shows the
results calculated from the \ion{Ne}{8}/\ion{O}{6} ratio.
The daily $A(Ne)/A(O)$ values fall in the range
0.08--0.16$\pm$0.01, also decreasing from the higher value in 2010 near solar minimum 
to the lower value near solar maximum in 2014, and 
the 27-day running average values fall from 0.13 in 2010 to 0.09 in 2014. The anti-correlation with the solar cycle is stronger 
at the higher temperatures associated with these spectral lines. The linear Pearson correlation coefficient is -0.68 for these data.

These results appear to indicate some fractionation between neon and oxygen over the solar cycle, despite both being 
high-FIP elements, as found in coronal streamers and the solar wind by \citet{landi&testa_2015} and \citet{shearer_etal2014}. In fact,
our results are in good agreement with the magnitude of the cyclic variations seen in the slow solar wind by \citet{shearer_etal2014}, who
found a fall from a high of 0.17$\pm$0.03 at solar cycle 23 minimum to 0.12$\pm$0.02 at the preceding solar cycle 23 maximum. A one-to-one comparison
is not possible since the measurements cover different cycle maxima, and their measurements are 6-month mean ratios, but there is some overlap. 
Our measurements start from 2010, April 30, which is in the
early ascending phase of the cycle and somewhat after solar minimum. Judging from their Figure 2, the value for the slow solar wind ($<$350\,km s$^{-1}$)
is around 0.14 at this time, in agreement with ours. This also suggets that the solar minimum value for the coronal $A(Ne)/A(O)$
ratio may be slightly higher than our measured maximum values i.e. our maximum values do not quite represent the truly unfractionated value.
We should note also that the 27-day Carrington averaged and 6 month mean data are less likely to give us a true measure at solar minimum. This is because
although days without sunspots are common near solar minimum, the Sun is less likely to be completely clear of spots for a whole month or half a year.

Since the $A(Ne)/A(O)$ ratio decreases with increasing activity, this would be compatible with an enhancement in the 
lower-FIP element oxygen compared to neon, which would be consistent with the cyclic variation of FIP-bias found in Sun-as-a-star 
spectra by \citet{brooks_etal2017}. Note that \citet{brooks_etal2017} found a strong correlation between
FIP-bias, measured from low- and high-FIP elements, and F10.7\,cm flux, but the anti-correlation we find 
between the $A(Ne)/A(O)$ ratios and the 27-day running averaged F10.7\,cm flux is slightly weaker during this period.
This suggests that the correlation found by \citet{brooks_etal2017} is driven
by the behaviour of the low-FIP elements with the solar cycle.

Conversely, the results are also compatible with a depletion of the higher-FIP element neon compared to oxygen. Although we consider
it less likely, this could, in principle, be evidence
of mass dependent fractionation due to gravitational settling of the heavier neon. \citet{raymond_etal1997} found evidence that plasma
within large coronal loops in the equatorial streamer belt had lower abundances than the plasma at the streamer edge, and suggested this
could be due to gravitaional settling as a result of a longer confinement time within closed magnetic field. \citet{feldman_etal1999} 
found other supporting evidence from measurements of line intensities as a function of height above the solar limb. In this picture, 
the Sun-as-a-star observations near solar minimum are dominated by emission from relatively shorter lived, and smaller, quiet Sun loops,
where gravitational settling is less likely to occur, whereas the emission near solar maximum is dominated by larger and longer lived
active region loops where gravitational settling leads to a depletion of neon relative to oxygen. It is worth pointing out that both a
depletion of neon relative to oxygen due to gravitational settling, and an enhancement of the lower FIP element oxygen, work in the same
direction to decrease the $A(Ne)/A(O)$ ratio. So it is possible that some combination of the two effects could be at play. 

\subsection{Summary}

In conclusion, we have used the very extensive EVE database to produce high quality daily averaged spectra covering the rise of
the solar cycle from 2010 to 2014. We then used these spectra to measure the $A(Ne)/A(O)$ ratio in two temperature ranges and examine any solar cyclic 
trend when the Sun is viewed as a star. This is a much larger sampling of the solar cycle than in previous work. 
Using the \ion{Ne}{5} and \ion{O}{4} lines, we find a relatively weak dependence on the solar cycle with a moderate anti-correlation 
in the Sun-as-a-star data. The measured $A(Ne)/A(O)$ value close to minimum is 0.17$\pm$0.01 (0.15 27-day average) and
the value close to maximum is 0.11$\pm$0.01 (0.13 27-day average). Using the higher temperature \ion{Ne}{8} and \ion{O}{6} lines we find a stronger anti-correlation 
and a clearer cyclic variation, with a value close to minimum of 0.16$\pm$0.01 (0.13 27-day average) and
a value close to maximum of 0.08$\pm$0.01 (0.09 27-day average). 

Given the uncertainties associated with the Li-like lines, we must consider the results from the \ion{Ne}{5}/\ion{O}{4} ratio to be more
robust. The fact that it is sampling a lower temperature region than the \ion{Ne}{8}/\ion{O}{6} ratio, where signatures of elemental 
fractionation are more difficult to detect, however, suggests that the hints of a cyclic dependence seen at those temperatures are 
consistent with the clearer detection in the \ion{Ne}{8}/\ion{O}{6} measurements. 
In either case, the values 
are broadly consistent with previous measurements in the quiet Sun \citep{young_2005}, and earlier satellite and rocket flight 
observations of the
Sun-as-a-star \citep{schmelz_etal2005}, and are also consistent with measurements in the slow solar wind \citep{shearer_etal2014}. 
Our study therefore supports earlier findings that the $A(Ne)/A(O)$ ratio
is too low to solve the ``solar abundance problem'' and that, given the uncertainties, it is difficult to reliably conclude
that the enhancement seen at solar minimum in coronal streamers by \citet{landi&testa_2015},
that may have helped, can be detected to a similar degree in Sun-as-a-star observations. We again, however, emphasize that our measurements
do not extend fully into the deepest solar minimum of 2008--2009.

%% ------------------------------------------------------------------------------------------
%% --- ACKNOWLEDGMENTS ----------------------------------------------------------------------
%% ------------------------------------------------------------------------------------------

\acknowledgments 

The work of DHB and HPW was performed under contract to the Naval Research Laboratory and was funded
by the NASA \textit{Hinode} program. DB and LvD-G are funded under STFC consolidated grant 
number ST/N000722/1. LvD-G acknowledges the Hungarian Research Grant OTKA K-109276.
The authors are grateful to Konkoly Observatory, Budapest,
Hungary, for hosting two workshops on Elemental Composition in Solar and Stellar Atmospheres
(IFIPWS-1, 13-15 Feb, 2017 and IFIPWS-2, 27 Feb-1 Mar, 2018) and acknowledge the financial 
support from the Hungarian Academy of Sciences under grant NKSZ 2018\_2. The workshops have
fostered collaboration by exploiting synergies in solar and stellar magnetic activity studies
and exchanging experience and knowledge in both research fields. CHIANTI is a collaborative
project involving George Mason University, the University of Michigan (USA) and the 
University of Cambridge (UK). The SDO data are courtesy of
the NASA/SDO, and the AIA, EVE, and HMI science teams. Hinode is a Japanese mission developed 
and launched by ISAS/JAXA, collaborating with NAOJ as a domestic partner, NASA and STFC (UK) 
as international partners. Scientific operation of the Hinode mission is conducted by the 
Hinode science team organized at ISAS/JAXA. This team mainly consists of scientists from 
institutes in the partner countries. Support for the post-launch operation is provided by 
JAXA and NAOJ (Japan), STFC (U.K.), NASA, ESA, and NSC (Norway).

{\it Facilities:} \facility{SDO (EVE), Hinode (XRT), SOHO (CDS, SUMER) }

%% ------------------------------------------------------------------------------------------
%% --- REFERENCES ---------------------------------------------------------------------------
%% ------------------------------------------------------------------------------------------

\end{document}